\documentclass[twocolumn,apj,floats,aps,amsmath,amssymb,nofootinbib]{emulateapj}
\usepackage{graphicx}
\usepackage{bm}
\usepackage{hyperref}
\usepackage[english]{babel}
\usepackage{tabu}
\usepackage{amsmath}

\begin{document}

\citestyle{egu}
\bibliographystyle{apj}

\title{The role of the cooling prescription for disk fragmentation:\\
Numerical convergence \& critical cooling parameter in self-gravitating disks}
\author{Hans Baehr\altaffilmark{1}}
\author{Hubert Klahr}
\affiliation{Max Planck Institute for Astronomy, K{\"o}nigstuhl 17, 69117 Heidelberg, Germany}
\altaffiltext{1}{baehr@mpia.de}

\begin{abstract}
Protoplanetary disks fragment due to gravitational instability when there is enough mass for self-gravitation, described by the Toomre parameter, and when heat can be lost at a rate comparable to the local dynamical timescale, described by $t_{c}=\beta\Omega^{-1}$. Simulations of self-gravitating disks show that the cooling parameter has a rough critical value at $\beta_{\textnormal{crit}}=3$. When below $\beta_{\textnormal{crit}}$, gas overdensities will contract under their own gravity and fragment into bound objects while otherwise maintaining a steady state of gravitoturbulence. However, previous studies of the critical cooling parameter have found dependence on simulation resolution, indicating that the simulation of self-gravitating protoplanetary disks is not so straightforward. In particular, the simplicity of the cooling timescale $t_{c}$ prevents fragments from being disrupted by pressure support as temperatures rise. We alter the cooling law so that the cooling timescale is dependent on local surface density fluctuations, a means of incorporating optical depth effects into the local cooling of an object. For lower resolution simulations, this results in a lower critical cooling parameter and a disk more stable to gravitational stresses suggesting the formation of large gas giants planets in large, cool disks is generally suppressed by more realistic cooling. At our highest resolution however, the model becomes unstable to fragmentation for cooling timescales up to $\beta = 10$.
\end{abstract}


\maketitle

\section{Introduction}
\label{sec:intro}

Between the two major theories of planet formation, core accretion and gravitational instability (GI), only the latter shows a tendency to form gas giant planets in wide orbits around young stars. The core accretion scenario, whereby the coagulation of dust particles forms larger objects like planetesimals and eventually cores, is the dominant mode of planet formation \citep{Janson2011}. It is particularly efficient at forming solid objects within $10$ au due to the amount of solid material and high collision rate \citep{Safronov1969,Pollack1996}. However, it is a slow process in outer regions, forming planetary cores with masses $>10 M_{\oplus}$ at timescales longer than the lifetime of the disk ($10^{6}$ to $10^{7}$ years), whereas the timescale for generating dense gaseous clumps by gravitational instability is significantly shorter at around $1000$ years \citep{Boss1997}.  At distances beyond around 30 au, protoplanetary disks become gravitationally unstable when they accrete enough mass such that regions within the disk collapse due to self-gravity, creating fragments that can accumulate solid material to form cores of Jovian planets or possibly form low mass stars \citep{Boss1997}. Thus discoveries of planets with large orbital radii \citep{Rameau2013,Marois2008} suggest that GI may be a feasible formation mechanism to form planets where core accretion has trouble operating, but not common enough to form planets at higher rates than via core accretion \citep{Janson2011,Janson2012}.

Gravitational instability becomes a significant factor in the disk when enough cold gas is present for strong self-gravitation and the disk cools efficiently \citep{Gammie2001,Durisen2007}. These two conditions are measured by the Toomre parameter $Q$ and a simple cooling relation $t_{c}$, respectively. The Toomre parameter represents a balance of the local centripetal and gravitational forces on a contracting fragment of the disk and is defined by \citet{Toomre1964} as:
\begin{equation} \label{eq:toomre}
Q = \frac{c_{s}\Omega}{\pi G\Sigma},
\end{equation}
where $\Omega = \sqrt{GM / R^{3}}$ is the orbital Keplerian frequency of the disk, which stabilizes the disk to large wavelength density perturbations and $c_{s}$ is the local speed of sound, which stabilizes the disk to shorter density perturbations \citep{Durisen2007}. Protoplanetary disks are generally considered thin, so the surface density is the vertically integrated density $\Sigma \approx \rho H$, where $H$ is the scale height of the disk. Regions of the disk with values below $Q \approx 1$ are unstable to axisymmetric perturbations and will contract into denser clumps, while a region with $Q > 1$ will remain stable to gravitational collapse \citep{Toomre1964}. When stable, a disk tends to stay close to $Q \approx 1$ as higher values mean there is more heat to be lost and the disk will cool faster and contract. Falling below $Q \approx 1$ results in strong shock heating which raises the temperature and returns the disk to marginal stability \citep{Balbus1999}. This gravitoturbulent situation arises when these effects balance each other and the simulation settles for several orbital periods.

When a fragment has formed it must remain bound under its own gravity and must remain cool enough so that a clump can collapse on a free-fall timescale shorter than shear can disrupt it \citep{Kratter2011}. Therefore a short cooling law
\begin{equation} \label{eq:cool}
t_{c} = \beta\Omega^{-1}
\end{equation}
is the second condition for creating an non-fragmenting disk, as a disk that can dump enough energy will form collapsed fragments which can withstand being torn apart by shocks and tidal shearing forces. Sufficient cooling has been an issue for the applicability of GI in the past \citep{Rafikov2005}, but it has been shown that short enough cooling times \citep{Tomley1991,Tomley1994}, represented in this paper by values below the critical value of $\beta_{\textnormal{crit}}=3$ \citep{Gammie2001}, it is possible to form gravitationally bound clumps that survive the shear of the disk.

\citet{Meru2011b} found $\beta_{\textnormal{crit}}$ changes with resolution, first for global smoothed particle hydrodynamic (SPH) simulations and later followed by local finite-difference shearing sheet models of \citet{Paardekooper2012}. These results would indicate that the critical cooling timescale may be significantly longer, up to $t_{c}=20\Omega^{-1}$, extending the GI formation domain to regions of the disk within 30 au (see Figure 2 of \citet{Janson2012}). This conflicts with our current understanding of formation regimes of core accretion and gravitational instability, suggesting that GI may be more common than detected by current observations \citep{Janson2011,Janson2012}. Additionally, this is a significant problem to the applicability of previous results as a resolution dependent solution implies that all previous studies at different resolutions produce different fragmentation conditions.

The widely used simple cooling law cools every location at the same rate, failing to account for the effects of optical depth on cooling efficiency or the increased strength of surface density fluctuations with resolution. To account for the varying optical depth with density, we alter the cooling to include a linear dependence on the local surface density in the disk. This is expected to suppress the formation of fragments from strong density fluctuations which will cool slower and be supported against collapse by higher internal temperature. Conversely, underdense regions will cool faster and prone to collapse, perhaps leading to convergence of the fragmentation boundary with resolution.

Other studies of thermodynamics on disk stability have focused on different methods, such as radiative cooling \citep{Boley2006,Cai2006} and modifications to the equation of state \citep{Laughlin1998} or have concentrated on different sources of heat such as irradiation from external sources \citep{Rice2011,Kratter2011,Vorobyov2015}. The closest approach to what is carried out here, \citet{Paardekooper2012}, showed with two-dimensional finite difference shearing sheet simulations, a simple cooling scheme does not lead to a converged critical cooling parameter for fragmentation. \citet{Rice2014} attempted to attain convergence by adjusting the numerical cooling and viscosity with a SPH code, but the critical cooling parameter identified therein does not agree with the results of both \citet{Gammie2001} and this paper.

The aim of this paper is to investigate the relation between the cooling law of the disk and the simulation resolution and their effect on formation conditions of gas giants planets. This means implementing a more realistic cooling relation per grid cell in local simulations, one dependent on the local surface density to mimic the change in optical depth for varying densities. This paper will proceed with an overview of the important equations in hydrodynamic simulations, the shearing sheet approximation and other theoretical considerations in Section \ref{sec:theory} followed by the specifics of the numerical setup in Section \ref{sec:numerical}. Finally, we look at the results of simulations and how they differ from previous attempts in Section \ref{sec:results} and discuss the limitations and implications in Section \ref{sec:discussion}.

\section{Theory}
\label{sec:theory}

To obtain high resolution simulations of disk dynamics the situation needs to be reduced to an numerically less expensive problem. Here we consider a fully gaseous disk with minimal magnetization, so the physical laws are the normal hydrodynamic (HD) equations. Since a very cool, thin disk is being considered, the equations and simulations here are only in radial and azimuthal directions. Like many astrophysical fluids, we treat the disk as a fluid with a high Reynolds number so we can describe its behavior in a disk around a young stellar object (YSO) using Eulers equation for the conservation of momentum and the typical equations for mass and energy conservation.
\begin{align}
& \frac{\partial\Sigma}{\partial t} + \nabla\cdot(\Sigma\mathbf{v}) = 0 \label{eq:massconserve} \\
& \frac{\partial\mathbf{v}}{\partial t} + (\mathbf{v}\cdot\nabla)\mathbf{v} = -\frac{\nabla P}{\Sigma} - \nabla\Phi \label{eq:momconserve} \\
& \frac{\partial\epsilon}{\partial t} + \nabla\cdot(\epsilon\mathbf{v}) = -P\nabla\cdot\mathbf{v}, \label{eq:energyconserve}
\end{align}

Since we are studying GI we consider the disk to be self-gravitating, defined by a potential $\Phi$ which is the solution to the Poisson equation for a razor-thin disk 
\begin{equation} \label{eq:poisson}
\nabla^{2}\Phi=4\pi G\Sigma\delta(z),
\end{equation}
which is solved in Fourier space by transforming the surface density to find the potential at the scale of wavenumber $k$ and transforming the solution back into real space. The solution to the Poisson equation in Fourier space is
\begin{equation} \label{eq:gravpotential}
\Phi(k_{x}, k_{y}, t) = -\frac{2\pi G\Sigma(k_{x}, k_{y}, t)}{| \mathbf{k} |}.
\end{equation}
In these simulations we smooth self-gravity on the grid scale and do not limit small wavelength modes in the calculation of self-gravity.

Finally, we consider the gas in the disk to be ideal, with surface density $\Sigma$, internal energy $\epsilon$, and 2D specific heat ratio $\gamma$
\begin{equation} \label{eq:eos}
P = (\gamma - 1)\Sigma\epsilon.
\end{equation}
The selection of the specific heat ratio has an effect on the cooling rate required for fragmentation \citep{Rice2005}. Higher values of $\gamma$ result in a stiffer equation of state that requires a lower cooling time for fragmentation. A ratio of $\gamma = 1.6$ is used here which compared to the value of $\gamma = 2$ by \citet{Gammie2001} might result in fragmentation at a slightly higher value of the critical cooling timescale $\beta_{\textnormal{crit}}$.

  \subsection{Shearing Sheet Model}
  \label{subsec:shear}

From these initial equations we move to a local description of a small section of the disk using a sheering sheet approximation. The disk is modeled locally on a small radial-azimuthal patch of the disk, transforming the global cylindrical coordinates to local Cartesian coordinates co-rotating with the disk \citep{Goldreich1965}. These assumptions allow for the modeling of the local properties of the disk while following the evolution of fragments that form when using periodic boundary conditions. This approximation ignores other properties of the disk such as accretion, non-local stresses and migration, which are saved for global simulations. Following this model, the relevant equations are similar to the conservation equations as above but with additional terms for the Coriolis effect $2\Omega\times\mathbf{u}$ and centripetal force $q\Omega v_{x}\mathbf{\hat{y}}$ as well as heating $H$ and cooling terms $C$ in the equation for conservation of energy. Additionally, the Pencil code used for these simulations uses entropy $s$ as the thermodynamic variable in the conservation of energy equation.
\begin{align}
\frac{\partial {\Sigma}}{\partial t} &- q\Omega x\frac{\partial {\Sigma}}{\partial y} + \nabla\cdot(\Sigma\mathbf{u}) = f_{D}(\Sigma) \label{eq:finalmassconserve} \\
\frac{\partial \mathbf{u}}{\partial t} &- q\Omega x\frac{\partial \mathbf{u}}{\partial y} + \mathbf{u}\cdot\nabla\mathbf{u} = \nonumber \\ 
& -\frac{\nabla P}{\Sigma} + q\Omega v_{x}\mathbf{\hat{y}} - 2\Omega\times\mathbf{u} - \nabla\Phi + f_{\nu}(\mathbf{u}) \label{eq:finalmomconserve} \\
\frac{\partial s}{\partial t} &-q\Omega x\frac{\partial s}{\partial y} + (\mathbf{u} \cdot \nabla)s = \nonumber \\
& \frac{1}{\Sigma T} \left( 2\Sigma\nu\mathbf{S}^{2} - \Lambda + f_{\chi}(s) \right) \label{eq:finalenergyconserve}
\end{align}
In these equations, $\mathbf{u} = (v_{x},v_{y}+q\Omega x )^{T}$ is the perturbed velocity in the disk due to the shear in the local box. Viscous heat is generated by $H = 2\Sigma\nu\mathbf{S}^{2}$, with rate-of-strain tensor $\mathbf{S}$, and radiated away by an approximation $\Lambda$ described in greater detail later in Section \ref{sec:theory}. The source terms $f_{D}(\Sigma)$, $f_{\nu}(\mathbf{u})$, $f_{\chi}(s)$ for hyperdiffusion, hyperviscosity, and hyperconductivity respectively are explicit terms to keep the solution well-behaved when shocks arise and will be expanded upon in Section \ref{sec:numerical}.

  \subsection{The $\alpha$-parameter}
  \label{subsec:alpha}

An important disk parameter to be determined by our shearing sheet model is the $\alpha$ stress, a means to measuring and comparing the sources of turbulence in a disk \citep{Shakura1973}. This formalism relies on the assumptions that the disk is thin and that angular momentum is transported locally through a dimensionally defined viscosity $\nu = \alpha c_{s} H$ \citep{Pringle1981}. A thin disk means that the only non-vanishing term of the vertically integrated stress tensor $\mathbf{T}$ is 
\begin{equation} \label{eq:stresstensor}
T_{r\phi} = -r\Sigma\nu\frac{d\Omega}{dr},
\end{equation}
which, when adding the above description of viscosity, becomes
\begin{equation} \label{eq:stresstensor2}
T_{r\phi} = -\alpha P\frac{d\ln\Omega}{d\ln r}.
\end{equation}
This shows the total stress in the disk comes only from the pressure and therefore viscosity is mostly local. Solving for $\alpha$ and defining the stress as the sum of the average gravitational and Reynolds stress tensors \citep{Cossins2009} gives:
\begin{equation} \label{eq:alpha}
\alpha = -\left(\frac{d\ln\Omega}{d\ln r}\right)^{-1}\frac{\langle G_{xy}\rangle + \langle H_{xy}\rangle}{\Sigma c_{s}^{2}},
\end{equation}
which allows us to calculate the $\alpha$ generated by the simulation, where
\begin{equation}
\langle G_{xy}\rangle = \int_{-\infty}^{\infty} \frac{g_{x} g_{y}}{4 \pi G} dz
\end{equation}
and
\begin{equation}
\langle H_{xy}\rangle = \langle \Sigma u_{x} u_{y} \rangle.
\end{equation}

The analytic expression of $\alpha$ based solely on input parameters $\gamma$ and $\beta$ is given by \citet{Gammie2001}:
\begin{equation} \label{eq:alphapara}
\alpha = \frac{4}{9}\frac{1}{\gamma(\gamma-1) t_{c}\Omega}
\end{equation}
which gives a prediction for the stress in a gravitoturbulent viscously heated disk. From equations (\ref{eq:alphapara}) and (\ref{eq:alpha}) we are able to compare theoretical expectations of the gravitoturbulent state with the stresses generated by our simulations.

  \subsection{Heating and Cooling}
  \label{subsec:cool}

For a disk to become unstable and its fragments to survive, it needs to cool fast enough to overcome the stabilizing effects of shocks and shearing motions. This means one needs a careful description of how the disk is treated thermodynamically; how and where heat is generated and released from the disk. The only active sources of heating come from viscous heating and shock dissipation, the first and third terms on the right side of equation (\ref{eq:finalenergyconserve}). Passive heating comes in the form of irradiation by the central star or other nearby stars, which limits the cooling from the surface.

A more physically accurate model currently in use is full radiative transfer, which typically requires three dimensional simulations with intricate opacities (see \citet{Boss2001}, \citet{Boley2006} and \citet{Cai2006}). Simulating cooling by radiative transfer is complicated, requiring significant computational resources with results differing on whether there is enough cooling to form fragments by gravitational instability. The effect of varying opacities with a simple cooling timescale has been investigated by \citet{Johnson2003} but without looking into the effect of varying resolutions and the stronger surface density perturbations which arise with increasing resolution.

To use computational resources more efficiently, simulations in this paper are run with a cooling law such that each grid cell loses an amount of heat per time given by the cooling law in the form $\Lambda=U/t_{c}$
\begin{equation} \label{eq:coolinglaw2}
\Lambda = \frac{\Sigma (c_{\textnormal{s}}^{2} - c_{\textnormal{s,irr}}^{2})}{(\gamma -1) t_{c}},
\end{equation}
where $U=\Sigma (c_{\textnormal{s}}^{2} - c_{\textnormal{s,irr}}^{2}) / (\gamma -1)$ is the two-dimensional energy density with a non-zero background irradiation term $c_{\textnormal{s,irr}}^{2}$. A disk with no irradiation will allow the gas to cool to a lower temperature and thus have less support from gravitational collapse.

The simple cooling time is derived by the assumption that the cooling timescale $t_{c}$ is proportional to the shearing or dynamical timescale $\Omega^{-1}$. This means that $\beta = t_{c}\Omega$ will be near unity to allow for dense clumps to collapse before tidal disruption. This assumes however, that surface density fluctuations are small compared to the initial value $\Sigma_{0}$, but density fluctuations will become larger with increased resolution, making a constant cooling parameter less viable with higher resolutions.

Here we re-examine this approximation, providing a rationale for a cooling timescale which varies with surface density for each grid cell, cooling denser regions slower and less dense regions faster, offering a more realistic approach to disk cooling. In this way, cooling is not only more appropriate for the physical system, it also scales according to stronger surface density fluctuations with increased resolution.

To approximate the heat lost by the disk through radiation, we assume heat is lost at a rate per unit surface area according to the Stefan-Boltmann law 
\begin{equation} \label{eq:boltzmann}
\Lambda(\Sigma,U,\Omega) = 2\sigma T_{e}^{4},
\end{equation}
with Stefan-Boltzmann constant $\sigma$ and effective disk temperature $T_{e}$. From the \citet{Hubeny1990} treatment of an radiative transfer in a optically thick disk by the diffusion approximation, one can express the effective surface temperature in terms of the midplane temperature $T_{c}$ and the Rosseland mean optical depth $\tau_{\textnormal{R}}$ of the intervening disk material
\begin{equation} \label{eq:hubeny}
T_{e}^{4}=\frac{8}{3}\frac{T_{c}^{4}}{\tau_{\textnormal{R}}}.
\end{equation}
This yields a heat loss relationship in terms of the midplane temperature and the passive heating due to irradiation from the star \citep{Rafikov2015}
\begin{equation}
\Lambda=\frac{16}{3}\frac{\sigma (T_{c}^{4} - T_{\textnormal{irr}}^{4})}{\tau_{\textnormal{R}}}.
\end{equation}
From this and the internal energy $U$ one can write the cooling timescale
\begin{equation} \label{eq:completecool}
t_{c} = \frac{U}{\Lambda} = \frac{3}{16}\frac{U\tau_{\textnormal{R}}}{\sigma (T_{c}^{4}- T_{\textnormal{irr}}^{4})} \approx \frac{3}{16}\frac{U\Sigma\kappa}{\sigma (T_{c}^{4} - T_{\textnormal{irr}}^{4})},
\end{equation}
where the optical depth is estimated in the optically thick regime by $\tau_{\textnormal{R}} \approx \Sigma\kappa$. This approximation produces an additional surface density dependence of the cooling timescale compared to the standard cooling prescription. Since we are considering the cool region of the disk where opacities are dominated by ice grains, $\kappa$ has no additional $\Sigma$ dependence only a dependence on temperature which we ignore for now \citep{Bell1993}.

Therefore, we approximate the cooling timescale in equation (\ref{eq:coolinglaw2}) by including a linear dependence on surface density
\begin{equation} \label{eq:newcoolingtime}
t_{c} = \beta\left(\Sigma / \Sigma_{0}\right)\Omega^{-1}.
\end{equation} 
This replaces equation (\ref{eq:cool}), which is specific to constant optical depths whereas equation (\ref{eq:newcoolingtime}) will take into consideration the changing optical depths in the disk due to local surface densities over the course of the simulation. Previous studies using the simple cooling law (\ref{eq:cool}), notably \citet{Gammie2001}, have found a critical value of $\beta\simeq 3$ where disks tend to fragment for $\beta$ values below this critical value and will not fragment above it. It is important to note that this critical value is found for simulations with $1024 \times 1024$ grid cells, and will differ with changing resolution, which is the focus of this investigation.

  \subsection{Fragmentation}
  \label{subsec:fragmentation}

Gravitational instability does not necessarily result in fragmentation; if $1\leq Q\leq 2$ the disk will be unstable to nonaxisymmetric perturbations, or spiral arms, that will transport angular momentum, but not collapse into fragments, a so-called gravitoturbulent state \citep{Bodenheimer2002,Armitage2011}. Additionally, clumps may become overdense only to fall apart and allow a disk to settle to a steady state. Therefore, fragments need to both be able to form and survive disruption for a few orbits \citep{Kratter2010}. A fragment survives when its density is above the Roche limit, where the self-gravity of the fragment is sufficient to keep from being sheared apart by the tidal forces of the protostar\citep{Shi2013,Chandrasekhar1963}
\begin{equation}
\rho_{\textnormal{Roche}}=3.5 \frac{M_{*}}{R^3},
\end{equation} 
where $M_{*}$ is the mass of the central star and $R$ is the radial distance between the fragment and star.

Since all presented simulations are local and take no consideration of any absolute central mass, one can use the Keplerian frequency $\Omega = \sqrt{GM / R^{3}}$ of the shearing box and the sound speed $c_{s}=H\Omega$ to formulate an expression for the Roche surface density in terms of simulation scale quantities and gravitational constant $G$
\begin{equation} \label{eq:rochedensity}
\Sigma_{\textnormal{Roche}}=7 \frac{c_{s}^2}{HG},
\end{equation} 
where $\Sigma_{\textnormal{Roche}}=2\rho_{\textnormal{Roche}} H$ \citep{Vorobyov2009a}. Fragments have formed when clump density is greater than the Roche surface density for more than a few cooling timescales $t_{\textnormal{c}}$.

\section{Numerical Methods}
\label{sec:numerical}

The local simulation of a disk is conveniently handled by a finite difference, partial differential equation solver for compressible MHD equations. Used for my investigations is the Pencil Code\footnote{http://pencil-code.nordita.org/} \citep{Brandenburg2001}, which is chosen for its high-order numerical scheme and its modularity. As a finite difference code the simulation domain is divided into a grid of cells where physical quantities are calculated and advanced in discrete time-steps. 

The length of the time-step is determined by the Courant criterion with Courant constant $C_{0} = 0.4$, which must be satisfied for convergence to be possible
\begin{equation} \label{eq:timestep}
\delta t = C_{0} min \left( \frac{\delta x}{|u_{x}|+c_{s}}, \frac{\delta y}{|u_{y}|+c_{s}} \right),
\end{equation}
which is calculated over the entire domain to calculate a single time-step to advance the entire system uniformly. 

For a two-dimensional simulation this means that the derivatives for each basic physical quantity is calculated in each grid cell with an upwinding scheme to eliminate spurious Nyquist signals 
\begin{multline} \label{eq:upwindedcalc}
f_{0}^{'} = \frac{-2f_{-3}+15f_{-2}-60f_{-1}+20f_{0}+30f_{1}-3f_{2}}{60\delta x} \\
- \frac{\delta x^{5} f^{(6)}}{60} = D^{(up,5)} + O(\delta x^{6}).
\end{multline}
The code then proceeds to the next time step determined by the Courant condition.

Shocks present problems in hydrodynamical simulations as discontinuities cannot be represented by high-order polynomials, leading to additional minimums and maximums. Therefore explicit dissipation terms are added to the conservation equations above to smooth out the waves so they do not hinder the performance of the simulation. These terms have two parts, one being a sixth-order hyper dissipation method and the second being a localized shock-capturing method, active in regions with large negative velocity divergences. The hyperdiffusion term looks like 
\begin{equation} \label{eq:hyperdiff}
f_{D}(\Sigma) = \zeta_{D}(\nabla^{6}\Sigma),
\end{equation}
with analogous forms for both hyperviscosity $f_{\nu}(\mathbf{u})$ and hyperconductivity $f_{\chi}(s)$. The shock-capturing portions for each dissipative term are
\begin{align}
& f_{D}(\Sigma) = \zeta_{D}(\nabla^{2} \Sigma + \nabla\ln\zeta_{D}\cdot\nabla\Sigma) \label{eq:shockdiff} \\
& f_{\nu}(\mathbf{u}) = \zeta_{\nu} (\nabla (\nabla\cdot\mathbf{u}) + (\nabla\ln\Sigma + \nabla\ln\zeta_{\nu})\nabla\cdot\mathbf{u}) \label{eq:shockvisc} \\
& f_{\chi}(s) = \zeta_{\chi}(\nabla^{2} s + \nabla\ln\zeta_{\chi}\cdot\nabla s), \label{eq:shockconduc}
\end{align}
where the $\zeta$ term for each is analogous to the following viscous example
\begin{equation} \label{eq:shockterm}
\zeta_{\nu} = \nu_{sh}\langle max_{3}[(-\nabla\cdot\mathbf{u})_{+}] \rangle [min(\delta x, \delta y, \delta z)]^{2}.
\end{equation}

\begin{figure*}[ft]
\centering
\includegraphics[width=1\textwidth]{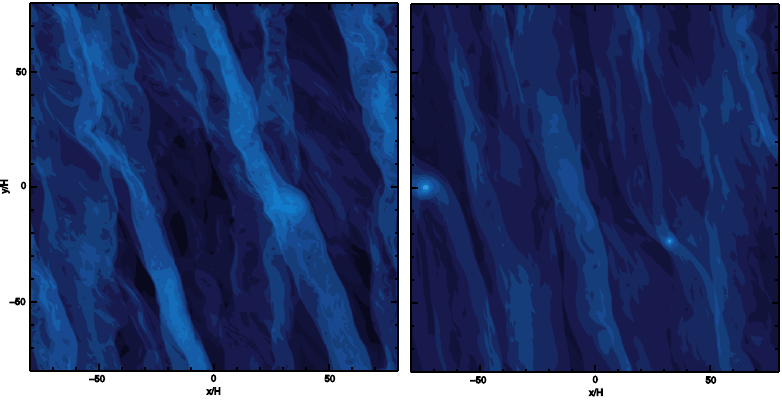}
\caption{Logarithmic surface densities of two different outcomes of self-gravitating disks. On the left is the stable gravitoturbulent state where heating and cooling are in balance and produce density waves. On the right is the case where cooling is short enough to allow an overdense clump to survive tidal shearing and become a bound fragment.}
\label{fig:densitymaps}
\end{figure*}

  \subsection{Boundary Conditions}
  \label{subsec:bound}

Typical of shearing sheet simulations, the boundary conditions are periodic along $y = 0$ and $y = L$ which means ghost zones are used to give conditions for grid cells near these boundaries \citep{Stone1992}. Derivatives are not calculated in these regions, only the values of density, velocity, etc. so that physical properties and their derivatives maybe calculated near the borders.
\begin{equation} \label{eq:ybounds}
f(x, y, t) = f(x, y + L, t)
\end{equation}
The boundaries $x = 0$ and $x = L$ require a different boundary condition on account of the shear velocity $u_{y} = v_{y} + \frac{3}{2}\Omega x$ \citep{Hawley1995}.
\begin{equation} \label{eq:xbounds}
f(x, y, t) = f(x + L, y - \frac{3}{2}\Omega Lt, t)
\end{equation}
When calculating this over a shear periodic x-boundary, the displacement due to the shear is taken into account by shifting the entire y-direction to make the x-direction periodic before proceeding with the transform in the x-direction. After the calculation of the potential in Fourier space the process is reversed to get back to real space.

\begin{figure}[t]
\centering
\includegraphics[width=0.5\textwidth]{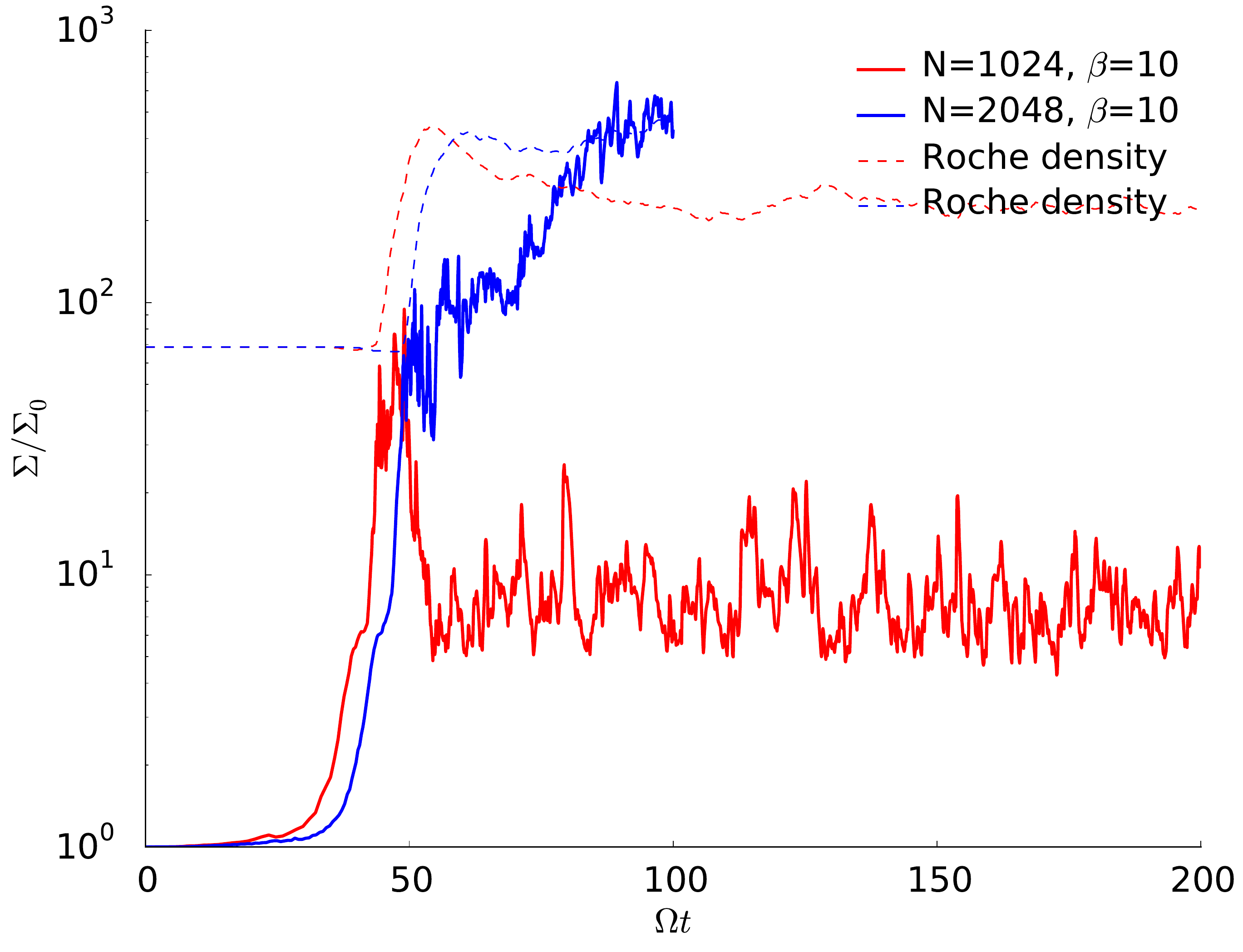}
\caption{A plot of maximum surface density over time shows the change in the fragmentation behavior two identical simulations aside from the resolution. The $N=2048$ case should not fragment if the critical cooling criterion is $\beta=3$ according to \citet{Gammie2001}.}
\label{fig:nonconverge}
\end{figure}

  \subsection{Initial Conditions}
  \label{subsec:init}

The Pencil Code uses dimensionless scale parameters for all physical values and constants. This helps keep numbers from getting too large or too small for the code to handle. The variables and constants can be scaled back to physical units after the computation is complete. The Keplerian frequency $\Omega = \sqrt{GM_{*}/R^{3}}$ is one of the more important parameters because it relates the shearing box to its surroundings, with $R$ the distance to the central massive object and $M_{*}$ being the mass of the central object.

The initial state of the disk is set so that the Toomre value throughout is on the borderline of stability and instability $Q=1$. By equation (\ref{eq:toomre}), this condition is met by setting the gravitational constant $G$, the Keplerian orbital frequency $\Omega$, and the uniform surface density distribution $\Sigma_{0} = 1$. Furthermore, the sound speed $c_{s}$ is initially set to $\pi$ which means the constant background irradiation set by $c_{s,0}^{2}$ in equation (\ref{eq:coolinglaw2}) is $\pi^{2}$. The physical length of the simulation needs to be longer than the critical wave length  $\lambda_{\textnormal{crit}}=2 c_{s}^2/G\Sigma_{0}$ \citep{Toomre1964} and so the physical length of the shearing box is $L_{x} = L_{y} = \mathbf{(160/\pi)} H$ for all runs, which also ensures a resolution of one scale height by at least 10 grid cells. The ratio of specific heats is set to the 2D adiabatic case $\gamma = 1.6$, which maps to a 3D adiabatic index of between 1.6 and 1.9 depending on the self-gravitation in the disk \citep{Johnson2003}.

\section{Results}
\label{sec:results}

Simulations of self-gravitating disks have used simple cooling (\ref{eq:cool}) as a useful starting point to model disk cooling, but as a simple mathematical relationship between the cooling and dynamic timescales it does not remain consistent for changing resolutions because it does not consider that optical depths vary in the disk. Naturally the general inclination of these simulations is to push for higher resolution with the hope that the fragmentation boundary will eventually level off at some higher resolution, but this expectation is not realistic. Surface density fluctuations will continue to increase but without a physically motivated solution there is no way to adequately scale with resolution.

Here we will show that our initial setup is consistent with previous results, e.g. the fragmentation boundary varies with resolution and this boundary is $\simeq 3$ for the smallest number of grid cells considered $N_{x}=N_{y}=1024$. Finally we will show that changing the way the disk cools according to Section \ref{sec:theory} results in a disk that fragments at shorter cooling timescales, without demonstrating convergence.

  \subsection{Previous Results}
  \label{subsec:prev}

Initial simulations run with $N_{x}=1024$ and $N_{y}=1024$ are established with a uniform surface density distribution, linear velocity shear and subsequently heated by shocks that develop from trailing density structures and cooled assuming large \textit{constant} optical depth. At this resolution all simulations were adjusted to be consistent with the fragmentation criterion $\beta_{\textnormal{crit} = 3}$, in which case a simulation with a cooling parameter of $\beta = 10$ does not fragment.  The simulation on the left of Figure \ref{fig:densitymaps} demonstrates the steady state of such a disk with non-axisymmetric density structures and roughly constant overall Toomre stability.

The behavior of the gravitoturbulent simulation is shown as the solid red line in Figure \ref{fig:nonconverge}, showing the maximum surface density to observe incidences of fragmentation. As discussed in \citet{Gammie2001}, initial small random velocities develop into non-linear fluctuations in surface density, velocity and gravitational potential before settling to a steady state in the non-fragmenting case, shown on the left of Figure \ref{fig:nonconverge} or continuing to grow in the fragmenting case, shown on the right of Figure \ref{fig:nonconverge}. As can be seen in the Figure \ref{fig:nonconverge}, in the stable gravitoturbulent case clumps are continuously forming but are torn apart before they are allowed to reach Roche density where they will be able to withstand further disruption. The trailing density structures lead to a finite $\alpha$ viscosity. By equation (\ref{eq:alphapara}), $\alpha=0.046$ for disks with cooling criterion $\beta=10$, and is observed in the stable case.

In the gravitationally unstable case (solid blue line of Figure \ref{fig:nonconverge}), fragments cool faster than they can be torn apart by tidal shear and collapse into one or more overdensities that may continue to grow or merge through collisions. Such behavior is expected for simulations with resolution $N=1024$ and cooling parameter $\beta \lesssim 3$, as a clump will be able to collapse to a compact density before being sheared apart over a few dynamical timescales. A disk is considered to fragment when it has surpassed the Roche surface density, indicated by the dashed lines for each simulation in Figure \ref{fig:nonconverge}, as defined in Section \ref{sec:theory}.

The difference between the two simulations shown in Figures \ref{fig:densitymaps} and \ref{fig:nonconverge} is that the total number of grid cells is quadrupled, which should not lead to such a drastic shift towards fragmentation. This is consistent with simulations by \citet{Meru2011b} and \citet{Paardekooper2012}, which indicate the critical cooling criterion of \citet{Gammie2001} may be as high as $\beta_{\textnormal{crit}} = 10$. Thus a constant cooling parameter does not adequately scale with resolution and a new approach to cooling is needed to observe a convergent fragmentation boundary.

  \subsection{Results with adjusted cooling}
  \label{subsec:new}

\begin{figure}[h]
\centering
\includegraphics[width=0.5\textwidth]{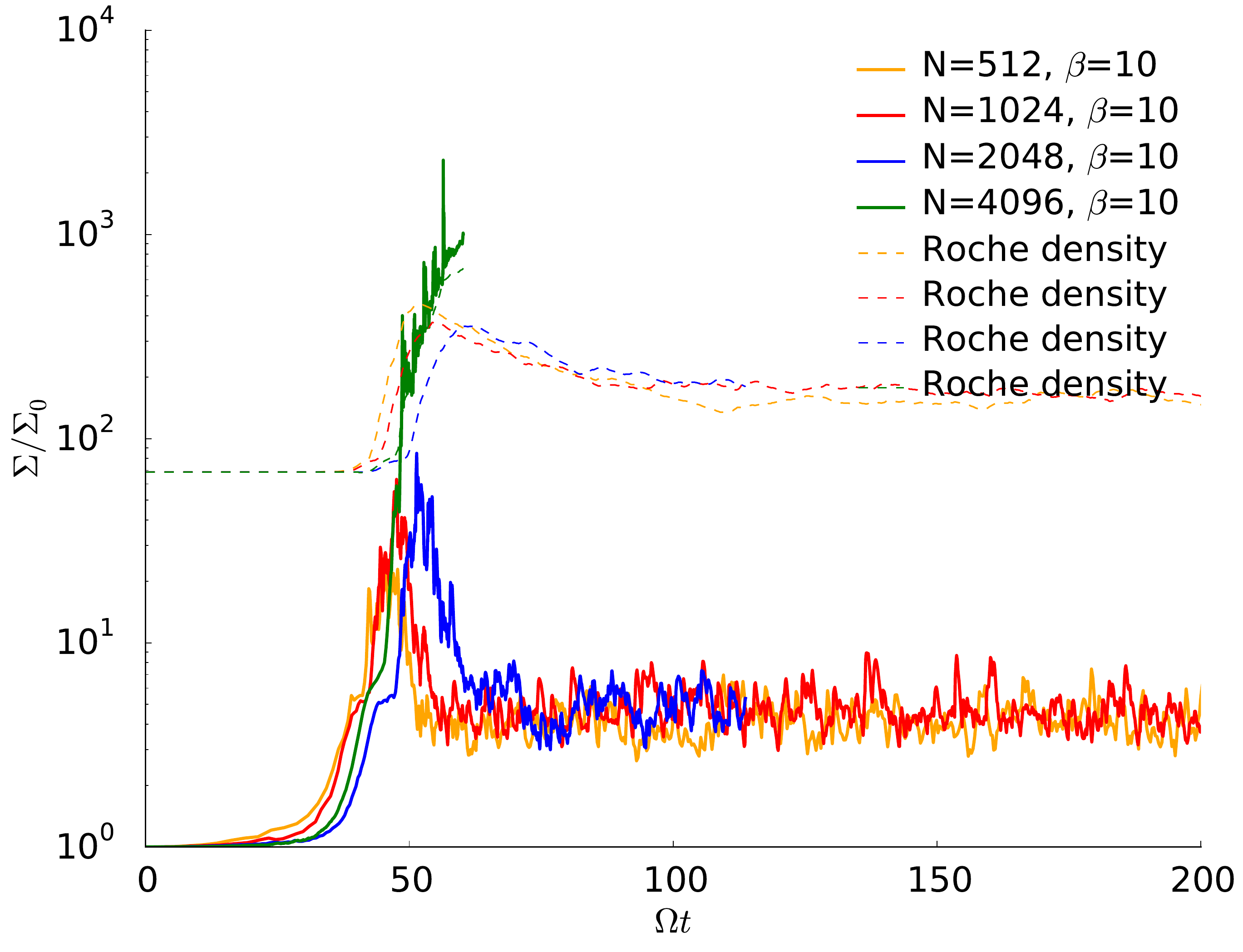}
\caption{The maximum surface densities of the two simulations shown in figure \ref{fig:nonconverge}, but with cooling prescription changed to account for the varying optical depths in the disk. The dashed lines are the corresponding Roche densities which are a fragmentation threshold.}
\label{fig:convergence}
\end{figure}

\begin{figure}[h]
\centering
\includegraphics[width=0.5\textwidth]{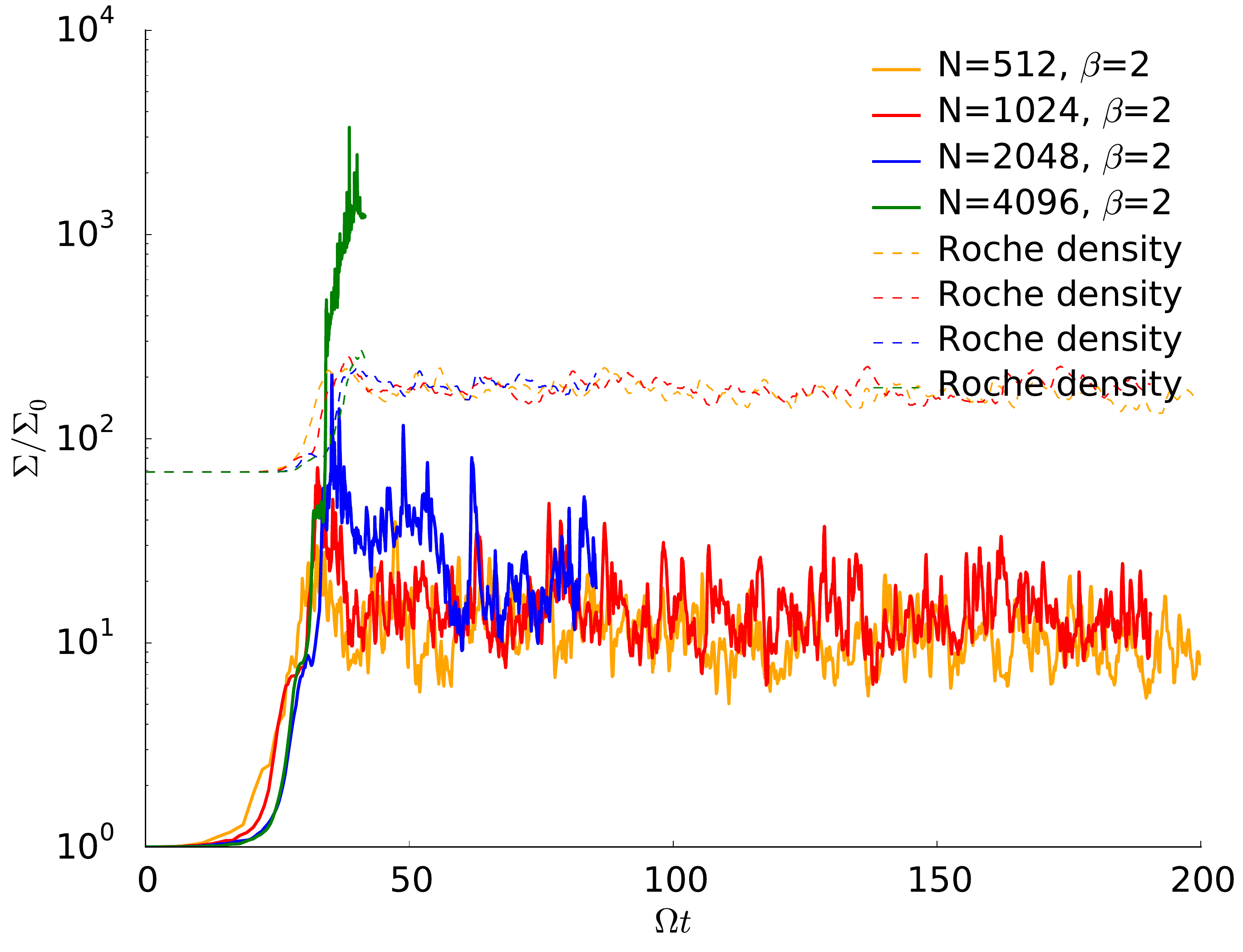}
\caption{Maximum density evolution of the three simulations from Table \ref{tab:newsims} with $\beta = 2$. Under the fragmentation criterion of \citet{Gammie2001} these simulations should fragment, but are instead gravitoturbulent with the change of cooling.}
\label{fig:converget2}
\end{figure}

The results here use parameters identical to the simulations in the previous section, besides the modification to the cooling law described in Section \ref{sec:theory} by equation (\ref{eq:newcoolingtime}). Figure \ref{fig:convergence} shows the case where $\beta = 10$ is shown for four different resolutions, all of which show non-fragmentation. These simulations are a direct comparison to the two shown in Figure \ref{fig:nonconverge} and in particular, the significant change between the behavior of the $N=1024$ and $N=2048$ cases shows the effect of altering the cooling prescription.

\begin{table}[h]
\caption{\textnormal{Simulations using the new cooling prescription.}}
\centering
\begin{tabular}{l*{6}{c}r}
Name  & Grid Cells ($N^2$) & $\beta$ & Fragmentation \\
\hline
hk10Q1               & $512^2$    & 10          & No \\
hk2Q1                 & $512^2$    & 2            & No  \\
hkp5Q1               & $512^2$    & 0.5         & No \\
1k10Q1               & $1024^2$  & 10          & No \\
1k5Q1                 & $1024^2$  & 2            & No \\
1kp5Q1               & $1024^2$  & 0.5         & No \\
2k10Q1               & $2048^2$  & 10          & No \\
2k5Q1                 & $2048^2$  & 2            & No \\
2kp5Q1               & $2048^2$  & 0.5         & Yes \\
4k10Q1               & $4096^2$  & 10          & Yes \\
4k2Q1                 & $4096^2$  & 2             & Yes \\
\end{tabular}
\label{tab:newsims}
\end{table}

This is expected because defining the cooling according to equation (\ref{eq:newcoolingtime}) creates an effective cooling time for each grid cell. When the disk cools according to the old cooling law $t_{c} = \beta \Omega^{-1}$, all regions in the disk lose the same amount of heat at the same timescale $t_{c}$. However, since one expects a clump to have a higher optical depth $\tau \approx \Sigma\kappa$ cooling efficiency should change from one location to another depending on the surface density. This motivates the alteration to the cooling timescale, which causes denser regions of the disk to have a higher optical depth and retain their heat, stabilizing to gravitational collapse. On the other hand, underdense regions will have a lower relative optical depth, cool faster and clump into dense structures easier.

\begin{figure*}[ft]
\centering
\includegraphics[width=0.75\textwidth]{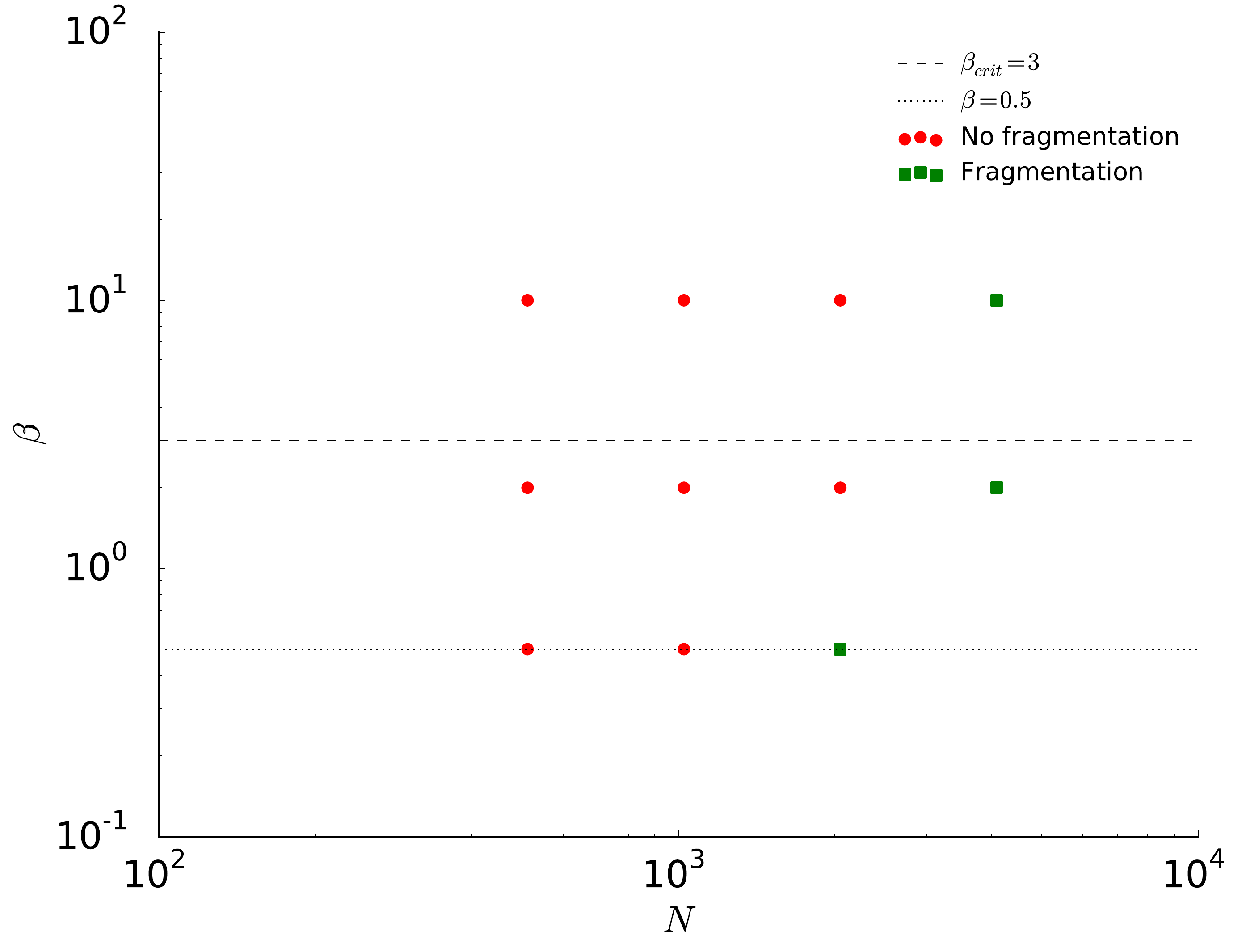}
\caption{Plot of cooling parameter $\beta$ against number of grid cells in one direction $N$ for all simulations run with the new cooling prescription from Table \ref{tab:newsims}. The dashed line at $\beta_{\textnormal{crit}} = 3$ is the fragmentation boundary as defined by \citet{Gammie2001}. The dotted line is the proposed new fragmentation boundary based on the simulations carried out here.}
\label{fig:convstudy}
\end{figure*}

\begin{figure}[h]
\centering
\includegraphics[width=0.5\textwidth]{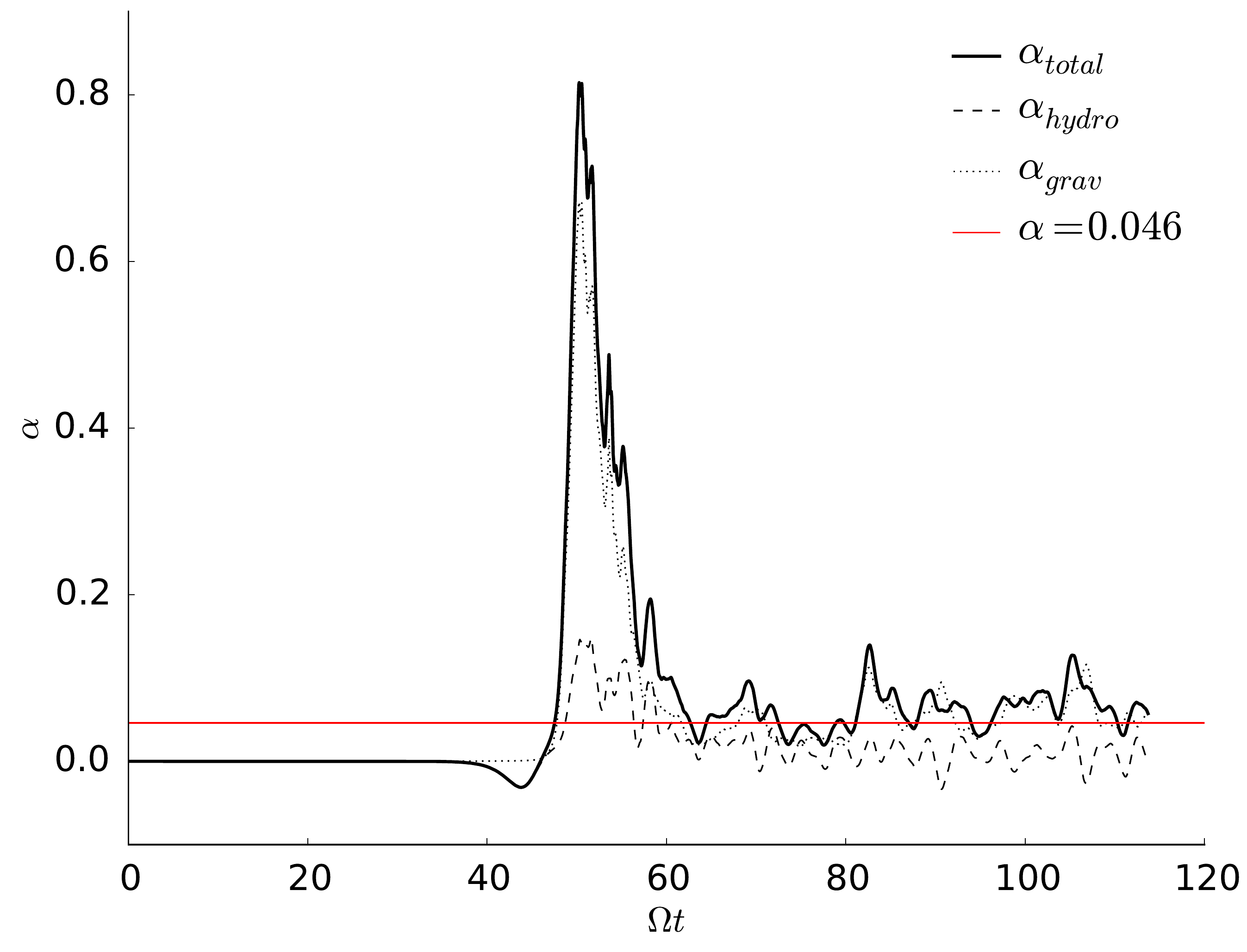}
\caption{The calculated (solid black line) versus predicted value (solid red line) of $\alpha$ for simulation named \textit{2k10Q1} which is a high resolution ($N=2048$) gravitoturbulent disk with adjusted cooling. Also plotted are the hydrodynamic and gravitational constituents of the $\alpha$ stress.}
\label{fig:alphadecomp}
\end{figure}

Figure \ref{fig:alphadecomp} shows the total $\alpha$ stress in a disk with adjusted cooling and the calculated value matches the expectation of the analytical prediction (\ref{eq:alphapara}). At early times, non-axisymmetric perturbations become unstable and contract causing a small drop in the total value of $\alpha$ just after $t_{\textnormal{c}} = 40\Omega^{-1}$. After this initial burst which is dominated by the shocks and self-gravity of the formed density structures, cooling takes over and the simulation settles to the expected gravitoturbulent $\alpha$-value.

The convergent behavior continues to lower cooling parameter values as well, Figure \ref{fig:converget2} showing the results of three simulations with different resolutions at $\beta =2$. Previously these simulations would have been expected to fragment, but here all show consistent steady gravitoturbulence. This begins to show the shift of the fragmentation boundary towards shorter cooling timescales. Figure \ref{fig:convstudy} shows the results of simulations at even short timescales with a new fragmentation boundary anticipated at around $\beta = 0.5$.

\section{Discussion}
\label{sec:discussion}

The results here show that the convergence issue of fragmentation in protoplanetary disks can not be approached by refining the numerical methods but by using more sophisticated physics. As simulations reach higher resolutions, they will likely need improved physical models to approach convergence. Additionally, the results here keep the formation regime of gravitational instability in the outer regions of disks where cooling times are sufficiently short. This is is contrary to other results which obtain convergence and see longer cooling timescale possible for the formation of planets, moving the formation region to shorter radii.

  \subsection{Convergence}
  \label{subsec:converge}
  
Whereas the old cooling timescale showed drastic differences in fragmentation behavior between two resolutions as seen in Figure \ref{fig:nonconverge}, the simulations here are consistent with each other over similar cooling timescales, cooling reaching a gravitoturbulent state at similar rates and settling at similar densities. This can be attributed to the sensitivity to density of the new cooling method employed here. When overdensities cool slower than other regions, they retain more heat and are more likely to be disrupted, decreasing the fragment density below the Roche density threshold and are suppressed from further fragmentation.

However this does not mean a new fragmentation boundary has been attained. Figure \ref{fig:convstudy} shows different fragmentation behavior for varying cooling times and resolutions and the lack of convergence is particularly noticeable in the cases where $\beta = 0.5$. The $N=1024$ simulation appears closer to fragmentation than the $N=512$ case and might be considered borderline fragmentation, where a clump surpasses the Roche density, but is sheared apart in less than an orbit ($2\pi\Omega^{-1}$) \citep{Meru2011b}. This is not the case for $N=512$ as a clump here never reaches Roche density and at the highest resolution studied ($N=2048$) the disk fragments. This may be due to the fact that assuming a simple linear relation in surface density does not fully capture the dependence of the cooling timescale on surface density. Also, as the cooling rate is a function of orbital frequency $\Omega$, surface density $\Sigma$ and temperature $T$ there may be an additional dependence on temperature that must be explored in the future.
  
\citet{Paardekooper2012} confirmed the non-convergence shown in SPH simulations with a finite difference code, which led to the assertion that fragmentation might be a stochastic process in circumstellar disks. The implication is that planet formation by GI is inevitable and the only reason that GI is not more prevalent is because the timescale for clumps in weakly cooling disks to achieve fragmentation is longer than the lifetime of the disk. This assumes that once fragmented, clumps do not fall apart and there is no process in the disk that could lead to the disruption of a successfully fragmenting clump. Our cooling implementation becomes weaker for a fragment as it increases in density, offering the necessary resistance to the stochastic formation of growing overdensities such that fragmentation is no longer an eventuality.

Consider the case of a clump which hovers very close to its Roche density, such as the solid blue line in figure \ref{fig:converget2}. In this case the simulation forms a clump which should have a better chance of crossing the threshold into becoming a fragment and remaining so. However, even as it manages to form a clump which crosses this threshold once, it still returns to a gravitoturbulent state at simulation time $t = 50\Omega^{-1}$, a result of the local cooling time. For this reason, these simulations do not indicate fragmentation is a strictly stochastic process independent on the strength of fluctuations.

\citet{Meru2011b} and \citet{Lodato2011} have suggested that the lack of convergence may be an effect of the numerical setups used, however \citet{Rice2014} did not find an issue with the artificial viscosity. \citet{Rice2014} does however alter they way in which an SPH method cools, using kernel smoothing to spread released heat around to neighboring particles. At high densities this has a similar effect to the cooling used here, with clusters of particles able to share their heat among each other so that dense clumps retain heat and resist collapse. At lower densities cooling is unchanged and this shows in their resulting critical cooling criterion which increases to $7 \leq \beta \leq 9$ compared to the reduction of the criterion in this study. Additionally, the ability of this cooling modification to remain consistent with particle number is uncertain, as it introduces a parameter, smoothing length, which should be scaled with resolution.

\citet{Rice2005} suggests that fragmentation is the result of the disk being unable to withstand the combined Reynolds and gravitational stresses which results in a fragmentation boundary at $\alpha\approx 0.1$. The simulations here do not support a fragmentation boundary at this value as some disks remain stable at values as high as $\alpha=1$. For simulations with the altered cooling time it is expected that the disk can remain stable to higher stresses because the localized cooling time stabilizes fragments and the disk as a whole.

  \subsection{Giant Planet Formation}
  \label{subsec:form}

The theories of how planets are formed are slowly starting to come to some agreement. Planetesimals formed by binary accretion of solid objects leads to the formation of rocky planets within 30 au of a young star with massive cores able to accumulate significant gaseous envelopes may become gas giant planets \citep{Mordasini2010}. This does not explain how to form large gas planets in outer regions of the disk, but disk instability offers a niche formation mechanism that can form planets in these regions given the right conditions. Since these distant gas planets are not very common and planet formation by instability is not an easy process, it appears that for now this is a reasonable explanation.

A suggestion by \citet{Meru2011b} is that the non-convergence of the critical cooling parameter could lead to large gas giants forming at longer cooling timescales, which implies shorter orbital radii. While the effect of shifting the inner boundary of gravitational instability would be minimal in the case of a single system, the effect on a large population could be more significant. If such a shift becomes more significant it could affect whether the formation regime of gravitational instability is in fact as restricted as observations seem to indicate.

The fragmentation boundary using the cooling scheme presented here shows a significant change compared to that of \citet{Gammie2001} and others. This means that the planet formation by disk instability is more restrictive than previously thought and certainly not heading in the direction of longer cooling timescales. This keeps the formation region of gravitational instability in a narrow region where the dominant theory for planet formation cannot form planets fast enough by the current understanding \citep{Janson2011,Janson2012}.

  \subsection{Limitations}
  \label{subsec:limits}
  
In this investigation, only a small modification has been made to the physics of the circumstellar disk and should not be expected to be a final solution to the convergence issue regarding gravitational instability. There are still some drawbacks to this approach though, as numerous assumptions and simplifications were made for the sake of efficient computation of high resolution physics and these might influence the evolution of the disk simulation as well as the fragmentation criteria.

As local simulations, these results do not take into account global parameters like accretion or long-range interactions, offering only a limited view of the disk. Therefore, these simulations do not take into consideration the chance that fragments may migrate or consider how the disk got to that state \citep{Kratter2011}. Fortunately, this does not seem to have a significant effect on the results of fragmentation criteria, with both local and global simulations in agreement on fragmentation criteria in general in its previous implementation \citep{Durisen2007}.

Radiative transfer is a more physically complete description of the cooling in the disk, but due to its relation to realistic opacities and the need for an additional dimension for effective simulation makes it a complicated option. Implementing radiative transfer in addition to adding a vertical computational direction significantly increases the amount of processing power needed for resolutions similar to what is implemented here. Using a simple cooling timescale in 3D leads to the same fragmentation behavior as \citet{Gammie2001} \citep{Mejia2005}, but including radiative transfer has not been shown to lead to consistent fragmentation, with cooling by radiative transfer too slow to form fragments as in \citet{Cai2006} and \citet{Boley2006}, but not in \citet{Boss2001}.

Questions have been raised about the razor-thin disk approximation and the calculation of self-gravity in such an approximation with \citet{Young2015} finding that the fragmentation boundary depends strongly on the gravitational smoothing used. Some studies of self-gravitating disks are smoothed on the grid scale (including those carried out here) which exaggerate the strength of self-gravity on small scales and may cause fragmentation at longer cooling timescales at higher resolutions and may be the cause of the non-convergence of the fragmentation boundary.

There are still some improvements which can be made to simulations which use simple cooling. A stronger dependence of the cooling timescale on surface density is possible due to an additional factor of $\Sigma$ in the energy density $U$ of equation (\ref{eq:completecool}). While opacity remains independent of surface density at low temperatures, the temperature dependence varies greatly at low temperatures \citep{Bell1993} and is difficult to model with this simple cooling prescription. For this reason we have only considered surface density in the cooling timescale and a more complete description would handle the changing temperature dependence of opacity.

\section{Conclusions}
\label{sec:conclusion}

The current understanding of planet formation is that core accretion is the dominant planet forming process. Core accretion shows the ability to form terrestrial and gaseous planets in a wide range of sizes in regions nearby the central star. But this mechanism does not explain the formation of a few gas giant planets that have formed at very large radii where core accretion takes far too long to occur before the gas in the disk is blown away. Thus, gravitational instability shows the ability to fill this niche by forming massive gas giant planets at radii beyond 50 au.

This picture of planet formation is generally well-formed, but recent results had suggested that it is not as clear as believed. Since gravitational instability showed to occur at shorter radii than expected, its formation region encroached on that of core accretion, possibly blurring the lines formation regions of the two mechanisms. We have carried out 2D hydrodynamic simulations of self-gravitating disks which show:  
\begin{itemize}
\item At low resolutions, a cooling timescale with surface density dependency results in a disk more stable to fragmentation by self-gravity, with a critical cooling timescale around $\beta \approx 0.5$. This means no clumps had a local cooling time short enough to overcome disruption from tidal shear.
\item The increased stability we find in our simulations suggests fragmentation preferentially in regions with short thermal relaxation times.
\item At our highest resolution however, simulations fragment even for long cooling timescales (up to $\beta = 10$) indicating that our approach with a surface density dependent cooling timescale did not result in convergence of the fragmentation boundary.
\item Many of the gravitoturbulent simulations are stable up to $\alpha=1$, with the gravitational stress component dominating the Reynolds component, and thus stable to very short cooling timescales.
\end{itemize}

For these reasons we have not found convergence of the fragmentation boundary by using a cooling timescale dependent on the surface density to mimic the effects of varying optical depth.

\begin{acknowledgments}

The authors thank the referee for their thorough and useful comments. We would like to thank the other members of the theory of planet and star formation group, in particular Andreas Schreiber, for their useful advice, patience and help with this project. We would also like to thank Ken Rice for his help in setting up the Pencil Code. Our simulations shown were run on the THEO cluster at the Rechenzentrum Garching (RZG) of the Max Planck Society and the JUQUEEN cluster of the J{\"u}lich Supercomputing Centre \citep{Stephan2015}.

\end{acknowledgments}

\bibliography{library}

\end{document}